\documentstyle[prl,aps,epsf,floats]{revtex}  
\newcommand\LQCD{{\Lambda_{\rm QCD}}}
\newcommand\sla[1]{#1\hskip-0.5em \slash} 
\newcommand\hq{{\hat q}}
\newcommand\hm{{\hat m}}
\newcommand\vev[1]{\langle{#1}\rangle}
\begin{document}

\twocolumn[
\hsize\textwidth\columnwidth\hsize
\csname@twocolumnfalse\endcsname

\title{Operator Product Expansion for Exclusive Decays: 
$B^+\to D_s^+e^+e^-$ and $B^+\to D_s^{*+}e^+e^-$}
\author{David H. Evans\footnotemark, Benjam\'{\i}n
Grinstein\footnote{bgrinstein@ucsd.edu}
 and Detlef R. Nolte\footnote{dnolte@ucsd.edu} \\[4pt]}
\address{\tighten{\it
Department of Physics,\\
University of California at San Diego, La Jolla, CA 92093 USA}\\[4pt]
UCSD/PTH 99--05\\[4pt] April 1999}

\maketitle

\tighten{
\begin{abstract}
The decays $B^+\to D_{s,d}^+e^+e^-$ and $B^+\to D_{s,d}^{*+}e^+e^-$
proceed through a weak and an electromagnetic interaction. This is a
typical ``long distance'' process, usually difficult to compute
systematically. We propose that over a large fraction of phase
space a combination of an operator product and heavy quark expansions
effectively turns this process into one in which the
weak and electromagnetic interactions occur through a local operator. 
Moreover, we use heavy quark spin symmetry to relate
all the local operators that appear in leading order of the operator
expansion to two basic ones. We use this operator expansion to
estimate the decay rates for $B^+\to D_{s,d}^{(*)+}e^+e^-$.

\end{abstract}
}
\vspace{0.2in}

]\narrowtext

\newpage
\addtocounter{footnote}{1}
\footnotetext{daevans@physics.ucsd.edu}\addtocounter{footnote}{1}
\footnotetext{bgrinstein@ucsd.edu}\addtocounter{footnote}{1}
\footnotetext{dnolte@physics.ucsd.edu}

$B$-mesons, with all their different decay channels, have become one
of the prime objects for checking the standard model and measuring its
parameters. This has led to the development and
ongoing planning of facilities and experiments dedicated to their
study. In the near future hadronic facilities will produce\cite{BTeV}
more than $10^{11}$ $B$-mesons per year, allowing studies of very rare
processes.  Only processes which afford a trustworthy quantitative
theoretical description can be used to verify the standard model.
There has been some interest recently in the decays $B^+ \to
D^{*+}_{d,s} \gamma$\cite{leb99} and $B^+ \to D^{(*)+}_{d,s}
e^+e^-$\cite{evans-99-1}.  However, it was already realized in
\cite{leb99,evans-99-1} that there were technical problems with the
calculations.  We demonstrate that over a large fraction of phase
space a combination of an operator product and heavy quark expansions
renders $B^+ \to D^{(*)+}_{d,s} e^+e^-$ computable. Our methods are
fairly general and can be applied to a variety of other processes.

The Operator Product Expansion (OPE) is commonly used in the
calculation of {\it inclusive} decay rates. One uses the optical
theorem and the performs an OPE on forward scattering amplitudes. We
will show that for a class of radiative decays one may use the OPE to
compute {\it exclusive} decay amplitudes.  There is a simple physical
motivation for the use of the OPE in the decay amplitude of, for
example, $B^+ \to D^{(*)+} e^+e^-$. Consider the hadronic part of the
amplitude to lowest order in electromagnetic and weak interactions,
\begin{equation}
\label{T-prod}
\langle D^{(*)+}| \int d^4x\;e^{iq\cdot x}
\; T(j^\mu_{\rm em}(x){\cal O}(0)) |B^+\rangle.
\end{equation}
Here ${\cal O}=\bar b\gamma^\nu(1-\gamma_5) u \bar
c\gamma^\nu(1-\gamma_5) d$ is a four quark operator responsible for
the weak transition, $j^\mu_{\rm em}$ is the electromagnetic current
and $T$ stands for time ordering of these operators. The energy
denominators from intermediate states of energy $E$ are $M_B-E$; since
the $B$ mass, $M_B$, is much larger than that of the $D$ meson and its
excitations,  many intermediate states contribute
significantly to the amplitude. Since the available
energy is much larger (in units of $\LQCD$) than the energy spacings
between intermediate states, the time ordered product in (\ref{T-prod})
should be well approximated by an expansion in local operators (OPE).

This OPE is not a short distance expansion since the momentum transfer
is not in the Euclidean domain. Hence its validity relies on
quark-hadron duality. This is exactly analogous to the the use of an
OPE in the computation of heavy hadron lifetimes. For large enough
$M_B$ violations to duality can be described by (uncomputable) powered
suppressed operators in the OPE\cite{chibisov}. Therefore, we can trust
at least the leading term in our computation of the amplitude.

To use an OPE in Eq.~(\ref{T-prod}) we  take the heavy quark
limit for the $b$ and $c$ quarks. Rather than the matrix element with
external physical mesons of Eq.~(\ref{T-prod}), we first consider a
Green function with two external quarks and two external
anti-quarks. For the momenta of the heavy quarks we write
$m_bv+k_b$ and $m_cv'+k_c$, while for the light quarks we take $k_u$
and $k_d$.  The residual momenta are small, $k_i\ll m_{b,c}$, provided
$w\equiv v\cdot v'$ remains of order unity. Just as in the case of
semileptonic inclusive decays\cite{chay} the combined heavy-quark and
operator product expansions give an expansion in powers of
$k_i/m_{b,c}$.

\begin{figure}
\centerline{
\epsfysize 3.5in
\epsfbox{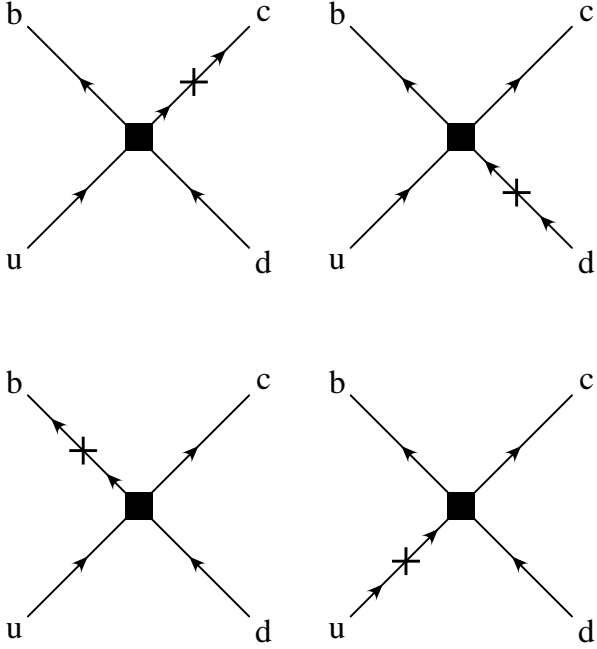}}
\vskip0.5cm
\caption{Feynman diagrams representing lowest order contributions to the Green
function. The filled square represents the four quark operator ${\cal
O}$ and the cross represents the electromagnetic current $j^\mu_{{\rm
em}}$, cf. Eq.~(\ref{T-prod}). }
\label{fig:fig1}
\end{figure}

To see how this works explicitly, consider the first  Feynman diagram of
Fig.~\ref{fig:fig1}. To lowest order in $k_i$ one has
\begin{equation}
\label{eq:feynm1}
Q_c\gamma^\mu\frac{m_b \sla{v}+m_c}{m_b^2-m_c^2}\gamma^\nu(1-\gamma_5)
\otimes \gamma_\nu(1-\gamma_5),
\end{equation}
where $Q_c=2/3$ is the charge of the $c$ quark. This corresponds to a
local operator which depends on the heavy masses and the velocity $v$,
which is not a kinematic variable but a parameter in the heavy quark
effective theory. Corrections appear as
higher dimension operators suppressed by powers of the large mass. For
example, the leading correction is of the form
\[
Q_c\gamma^\mu\frac{\sla{k}_c-\frac{2m_bv\cdot
k_c}{m_b^2-m_c^2}(m_b \sla{v}+m_c)}{m_b^2-m_c^2}
\gamma^\nu(1-\gamma_5)
\otimes \gamma_\nu(1-\gamma_5).\nonumber
\]
Of course, in the operator language, $k_c$ becomes $-i\partial$ acting
on the $c$-quark field.

Consider next the second Feynman diagram of
Fig.~\ref{fig:fig1}. Neglecting the light quark masses we have
\begin{equation}
\label{eq:feynm2}
-Q_d\gamma^\nu(1-\gamma_5)\frac{m_b\sla{v}-m_c\sla{v}'}{(m_bv-m_cv')^2}
\gamma^\mu
\otimes \gamma_\nu(1-\gamma_5).
\end{equation}
This term differs in an important way from~(\ref{eq:feynm1}). The
large denominator is not uniformly large over the whole kinematic
range. In fact the denominator is just the square of the leading part
of the momentum out of the electromagnetic current, $q=m_bv-m_cv'
+\sum k_i$. For the decay $B^+\to D^{(*)+}e^+e^-$, the kinematic range
is $0\le q^2\le (M_B-M_D)^2$. The approximation is valid
provided $\LQCD\ll m_{c,b}$, {\it e.g.,\/} the corrections are order $\LQCD/
m_{c,b}$. There are also corrections of order $\LQCD m_{b,c}/q^2$. So
our results are limited to the region were $q^2$ scales like
$m_{c,b}^2$. The region were $q^2$ does not scale like $m_{c,b}^2$ is
parametrically small, so the arguments we present are theoretically
sound. We emphasize that our
method cannot be applied to $B^+\to D^+\gamma$, for which $q^2=0$,
identically.

The third and fourth diagrams in Fig.~\ref{fig:fig1} are similarly
computed. In an obvious notation, these are
\begin{equation}
\label{eq:feynm3}
-Q_b\gamma_\nu(1-\gamma_5)\otimes 
\gamma^\mu\frac{m_b-m_c \sla{v}'}{m_b^2-m_c^2}\gamma^\nu(1-\gamma_5)
\end{equation}
and
\begin{equation}
\label{eq:feynm4}
-Q_u\gamma_\nu(1-\gamma_5)\otimes 
\gamma^\nu(1-\gamma_5)\frac{m_b\sla{v}-m_c\sla{v}'}{(m_bv-m_cv')^2}
\gamma^\mu.
\end{equation}
Finally, for  the calculation of the rate $B^+\to D^{(*)+}e^+e^-$
one must compute the matrix elements of these four local operators.  

The matrix elements of all of these local operators can be expressed,
by use of heavy quark spin symmetry, in terms of only two invariant
functions. 
Spin symmetry is best exploited using the Wigner-Eckart theorem. We
represent the meson by
\begin{equation}
H^{(c)}_{v'}=\left(\frac{1+\sla v'}{2}\right)\left[D^*_\nu\gamma^\nu
-D\gamma_5\right]
\end{equation}
and the anti-meson by
\begin{equation}
H^{(\bar b)}_{v}=
\left[B^*_\nu\gamma^\nu
-B\gamma_5\right]\left(\frac{1-\sla v}{2}\right).
\end{equation}
We will also need the field conjugate,
\begin{equation}
\bar H^{(c)}_{v'}=\gamma^0H^{(c)\dagger}_{v'}\gamma^0
=\left[D^{*\dagger}_\nu\gamma^\nu+D^\dagger\gamma_5\right]
\left(\frac{1+\sla v'}{2}\right).
\end{equation}
In terms of these we find
\begin{eqnarray}
\label{eq:beta-defd}
& &\langle H^{(c)}_{v'}| \bar h^{(c)}_{v'}\Gamma_cd\;
\bar h^{(\bar b)}_{v}\Gamma_bu |H^{(\bar b)}_{v}\rangle =\nonumber\\
& &\frac{\beta(w)}4{\rm Tr}({\bar H}^{(c)}_{v'}\Gamma_c) 
{\rm Tr}(H^{(\bar b)}_{v}\Gamma_b)
+\frac{\gamma(w)}{4}{\rm Tr}({\bar H}^{(c)}_{v'}\Gamma_c H^{(\bar b)}_{v}\Gamma_b),\nonumber\\
\end{eqnarray}
where $\Gamma_{b,c}$ are arbitrary $4\times4$ matrices, $\bar
h^{(c)}_{v'}$ is the field that creates a heavy quark with velocity
$v'$ and $\bar h^{(\bar b)}_{v}$ annihilates a heavy anti-quark with
velocity $v$. Invariance under spin symmetry of the heavy
quarks readily implies
\begin{equation}
\label{eq:fromspin2}
 \vev{H^{(\bar c)}_{v'}|\bar h^{(c)}_{v'}\Gamma_cd\;
\bar h^{(\bar b)}_{v}\Gamma_bu|H^{(b)}_v} \propto
 {\bar H}^{(c)}_{v'}\Gamma_c\otimes H^{(\bar b)}_{v}\Gamma_b.
\end{equation}
Invariance under rotations implies that the remaining four
indices must be contracted. 

The functions $\beta$ and $\gamma$ can be determined in simulations of
QCD on the lattice. In the calculations below they are estimated
using a vacuum insertion
approximation, which gives $\beta(w)=f_Bf_D\sqrt{M_BM_D}$ and
$\gamma(w)=0$. A better estimate can be obtained by
using isospin and heavy flavor symmetry to relate~(\ref{eq:beta-defd})
at $w=1$ to
\begin{eqnarray}
\vev{B|\bar b \gamma^\mu(1-\gamma_5) d\bar b \gamma_\mu(1-\gamma_5)
d|\bar B}
&=& \mbox{$\frac83$}B_Bf_B^2M_B^2,\\
 \vev{B|\bar b (1-\gamma_5) d\bar b (1-\gamma_5)
d|\bar B}
&=&- \mbox{$\frac53$}B_Sf_B^2M_B^2.
\end{eqnarray}
Isospin relates these matrix elements to the matrix element
in~(\ref{eq:beta-defd}) symmetrized in $u\leftrightarrow d$. This
symmetrized operator can be related\cite{evans-99-1}
to~(\ref{eq:beta-defd}) if we make the following assumption:
\begin{equation}
\vev{H^{(\bar c)}_{v'}|\bar h^{(c)}_{v'}\Gamma_cT^A d\;
\bar h^{(\bar b)}_{v}\Gamma_b T^Au|H^{(b)}_v}
=0.
\end{equation}
Here $T^A$ is a generator of color-$SU(3)$. This assumption is weaker
than vacuum insertion because it holds true even if we insert a
complete set of physical states, rather than only the vacuum, between
the currents.  We obtain $\beta(1)=\frac16(B_B+5
B_S)f_Bf_D\sqrt{M_BM_D}$ and
$\gamma(1)=-\frac56(B_B-B_S)f_Bf_D\sqrt{M_BM_D}$.  A recent
calculation\cite{GBS} using quenched lattice QCD gives
$B_B\approx B_S\approx0.8$, with a strange light quark and a $c$ and
$b$ heavy quark for $B_B$ and $B_S$, respectively. Therefore our
calculations below, which uses the vacuum insertion values
$\beta(1)=f_Bf_D\sqrt{M_BM_D}$ and $\gamma(1)=0$ can be trivially modified to account for
these lattice results, $\beta(1)\approx0.8f_Bf_D\sqrt{M_BM_D}$ and
$\gamma(1)\approx0$. We do not use these lattice results below because
$B_B$ and $B_S$ are computed for different heavy quarks and different
renormalizations, and the result for $B_B$ differs by 20\% from other
lattice studies\cite{latticeBB}.

The effective Hamiltonian for the weak transition is
\begin{equation}
{\cal H}'_{\rm eff}= \frac{G_F}{\sqrt2}\,V_{ub}V^*_{cd}\left(
c(\mu/M_W){\cal O}+c_8(\mu/M_W){\cal O}_8\right),
\end{equation}
where ${\cal O}_8=\bar b\gamma^\nu(1-\gamma_5)T^A u \bar
c\gamma^\nu(1-\gamma_5) T^A d$.  The dependence on the renormalization
point $\mu$ of the short distance coefficients $c$ and $c_8$ cancels
the $\mu$-dependence of operators, so matrix elements of the effective
Hamiltonian are $\mu$-independent. Resuming the leading logs, 
$c(\mu_0)=\frac13x^2+\frac23x^{-1}$ and $c_8(\mu_0)=x^{-1}-x^2$, where
$x=\left(\alpha_s(\mu_0)/\alpha_s(M_W)\right)^{6/23}$.

Defining
\begin{equation}
h^{(*)\mu}=
\langle D^{(*)+}| \int d^4x\;e^{iq\cdot x}
\; T(j^\mu_{\rm em}(x){\cal H}_{\rm eff}(0)) |B^+\rangle,
\end{equation}
the decay rate for $B^+\to D^{(*)+}e^+e^-$ is given in terms of $q^2$
and $t\equiv(p_D+p_{e^+})^2=(p_B-p_{e^-})^2$ by   
\begin{equation}
\label{eq:doublediffrate}
\frac{d\Gamma}{dq^2dt}=\frac1{2^8\pi^3M_B^3}
\left|\frac{e^2}{q^2}\ell_\mu h^{(*)\mu} \right|^2
\end{equation}
where $\ell^\mu=\bar u(p_{e^-})\gamma^\mu v(p_{e^+})$ is the leptons'
electromagnetic current. A sum over final state lepton helicities, and
polarizations in the $D^*$ case, is implicit. Using the OPE and spin
symmetry we obtain
\begin{eqnarray}
h^\mu=\frac\kappa3& &\Big[
\frac{(4wm_b-3m_c)v^\mu-(3m_b-2wm_c)v^{\prime\mu}}{(m_bv-m_cv')^2}
\nonumber\\
& &\hspace{3cm}-\frac{m_bv^{\prime\mu}+m_cv^\mu}{m_b^2-m_c^2}\Big]
\end{eqnarray}
and
\begin{eqnarray}
h^{*\mu}&=&\frac\kappa3\Big[\frac{m_b(3\epsilon^\mu
-4v\cdot\epsilon v^{\mu})+m_c(v\cdot\epsilon
v^{\prime\mu}-3w\epsilon^\mu)}{(m_bv-m_cv')^2}\nonumber\\
& &\hspace{2cm}
-\frac{3im_c\epsilon^{\mu\alpha\beta\gamma}\epsilon_\alpha v'_\beta v_\gamma}%
{(m_bv-m_cv')^2}
\\
&+& \frac{m_b\epsilon^\mu+m_c(v\cdot\epsilon
v^{\prime\mu}-w\epsilon^\mu)
+im_c\epsilon^{\mu\alpha\beta\gamma}\epsilon_\alpha v'_\beta v_\gamma}%
{m_b^2-m_c^2}
\Big].\nonumber
\end{eqnarray}
Here $\kappa=G_F/\sqrt2\,V_{ub}V_{cq}^*[c\beta+c_8\beta_8]$ and we
have defined the matrix element of ${\cal O}_8$ as $\beta_8$, in
analogy to Eq.~(\ref{eq:beta-defd}). These expressions are our central
results, demonstrating that the decay rates for $B^+\to
D^{(*)+}e^+e^-$ can be expressed in terms of the matrix elements
$\beta$ and $\beta_8$. Below we make an educated guess of these matrix
elements, but for reliable results they should be determined from
first principles, say, by monte carlo simulations of lattice QCD. 

There are short distance QCD corrections to the
coefficients~(\ref{eq:feynm1})--(\ref{eq:feynm4}) in the OPE. In
leading-log order they are determined by the renormalization
group. The matrix elements of the operators are all related by
spin symmetry to the reduced matrix elements $\beta$, $\beta_8$,
$\gamma$ and $\gamma_8$. However, $\beta$ and $\beta_8$ do not mix
into $\gamma$ and $\gamma_8$. Moreover, since we use $\gamma=\gamma_8=0$,
 we need only the $2\times2$ matrix
of anomalous dimensions of $\beta$ and $\beta_8$. 
A straightforward one loop computation gives
the anomalous dimension for $\beta$ and $\beta_8$ as follows:
\begin{equation}
\gamma=\frac{\alpha_s}{4\pi}
\left(\begin{array}{cc}
8 & 4(wr(w)-1)\\
\frac89 (wr(w)-1) & \frac43(7-wr(w))\\
\end{array}\right),
\end{equation}
where $r(w)\equiv\frac1{\sqrt{w^2-1}}\ln(w+\sqrt{w^2-1})$. Note that
the diagonal entry for $\beta$ is precisely twice the anomalous
dimension of the heavy-light current, as if the operator
was factorized. This is in accordance with the results of calculations of
$\gamma$ of the operator for $B-\bar B$ mixing\cite{vswp}. Moreover,
at $w=1$ the matrix simplifies, $\gamma=\frac{2\alpha_s}\pi{\bf 1}$.

Denote by $\tilde c$ and $\tilde c_8$ the coefficients of $\beta$ and
$\beta_8$, respectively. They
satisfy a renormalization group equation,
\begin{equation}
\label{eq:rge}
\mu\frac{d}{d\mu}{\bf \tilde c}=-\gamma^T{\bf \tilde c}.
\end{equation}
Here ``$T$'' denotes transpose of a matrix and ${\bf \tilde c}$ is a
column vector, ${\bf \tilde c}^T=(\tilde c, \tilde c_8)$. In
leading-log order the solutions are 
\begin{eqnarray}
\tilde c(\mu) &=&z^\psi[  \frac13(2{z}^{\xi}+{z}^{-\xi})\tilde c(\mu_0)
 +\frac29\left ({z}^{\xi}-{z}^{-\xi}\right ){\tilde c_8(\mu_0)}] \nonumber\\
\tilde c_8(\mu) &=&z^\psi[  \left ({z}^{\xi}-{z}^{-\xi}\right )\tilde c(\mu_0)
+\frac13\left ({z}^{\xi}+2{z}^{-\xi}\right )\tilde c_8(\mu_0)]\nonumber
\end{eqnarray}
where $z={\alpha_s(\mu)}/{\alpha_s(\mu_0)}$,
\begin{equation}
\psi=\frac{13-wr(w)}{3b_0}\quad,\quad
\xi=\frac{wr(w)-1}{ b_0}.
\end{equation}
$b_o$ is the coefficient of the one loop beta function in QCD,
$b_0=11-\frac23n_f$, with $n_f=3$ light flavors in our case. For our
numerical estimates below we match the coefficients $\tilde c$ and
$\tilde c_8$ to $c$ and $c_8$ at the intermediate scale
$\mu_0=\sqrt{m_bm_c}$.  Numerically, with
$\alpha_s(\mu_0)/\alpha_s(M_W)\approx2.24$ one has $c\approx1.0$ and
$c_8\approx-0.7$.  We will use $\mu=1~{\rm GeV}$, with the implicit
understanding that the matrix elements $\beta(w)$ and $\beta_8(w)$ are
computed at that value of the renormalization point.

It is now a trivial exercise to compute the differential decay
rate. We integrate the rate in Eq.~(\ref{eq:doublediffrate}) over the
variable $t$ and obtain, for $B^+\to D^{(*)+}_se^+e^-$,
\begin{equation}
\frac{d\Gamma}{dq^2}=
\frac{\alpha^2 G_F^2}{288\pi M_B^3}|V_{ub} V_{cs}|^2 
(c\beta+c_8\beta_8)^2{\cal F}(\hq).
\end{equation}
Here ${\cal F}(\hq)$ is a dimensionless function of $\hq \equiv
\sqrt{q^2/m_b^2}$ and $\hm\equiv M_{D_s}/M_B$. For $B^+\to D^+_se^+e^-$ it
is given by
\begin{equation}
{\cal F}(\hat q)=
\frac{4(2-{{\hat m}}^{2})^2[(1-(\hq+\hm)^2)(1-(\hq-\hm)^2)]^{3/2}}{3\hq^4\hm(1-\hm^2)^2}
\end{equation}
and for $B^+\to D^{*+}_se^+e^-$ by
\begin{eqnarray}
{\cal F}(\hat q)& &=
\frac43\frac{\sqrt{(1-(\hq+\hm)^2)(1-(\hq-\hm)^2)}}{\hq^6\hm(1-\hm^2)^2}
\nonumber\\
& &(-36 {\hm}^{8}+{\hm}^{2}\hq^{8}+9 {\hm}^{10}-\hq^{6}{\hm}^{4}+8 
{\hm}^{6}\hq^{4}\nonumber\\
& &-17 {\hm}^{8}\hq^{2}-30 \hq^{4}{\hm}^{4}+38 \hq^{2}{\hm}^{6}-4 \hq^{6}{\hm}^{2}+4 \hq^{6}\nonumber\\
& &+4 \hq^{2}-8 \hq^{4}-36 {\hm}^{4}+9 {\hm}^{2}-4 \hq^{2}{\hm}^{2}\nonumber\\
& &
+30 \hq^{4}{\hm}^{2}-21 \hq^{2}{\hm}^{4}+54 {\hm}^{6} ).
\end{eqnarray}
Note that we have not distinguished between heavy quark and meson
masses. The distinction enters at order $1/m_Q$ in the heavy mass
expansion, and we have not considered such corrections in this work. 

To estimate the branching fraction numerically, we use the vacuum
insertion approximation, $\beta(w)=z^{-4/b_0}f_Bf_D\sqrt{M_BM_D}$ and
$\beta_8(w)=0$, and integrate the rate from $q^2=1~{\rm GeV}$ up to
the kinematic limit $q^2=(M_B-M_{D_s})^2$. The lower limit is an
estimate of how low $q^2$ may be before our operator expansion breaks
down.  We find
\begin{eqnarray}
{\rm Br}(B^+\to D^{*+}_se^+e^-)|_{q^2>1~{\rm GeV}}
&=&1.8\times10^{-9}\\
{\rm Br}(B^+\to D^{+}_se^+e^-)|_{q^2>1~{\rm GeV}}
&=&2.7\times10^{-10}\\
{\rm Br}(B^+\to D^{*+}e^+e^-)|_{q^2>1~{\rm GeV}}
&=&9.1\times10^{-11}\\
{\rm Br}(B^+\to D^{+}e^+e^-)|_{q^2>1~{\rm GeV}}
&=&1.4\times10^{-11}
\end{eqnarray}
where we have used $|V_{ub}V_{cs}|=0.004$,
$|V_{ub}V_{cd}|=8.8\times10^{-4}$, $f_B=170~{\rm MeV}$ and
$f_D=f_B\sqrt{M_B/M_D}$. It is important to observe that the portion
of phase space $q^2>1~{\rm GeV}$ is expected to give a small fraction
of the total rate since the the pole at $q^2=0$ dramatically amplifies
the rate for small $q^2$.

In summary, we have applied the operator product and heavy quark
expansions to the exclusive decay amplitude of heavy mesons. At zero
hadronic recoil the matrix element of the leading operator in the OPE
is related by heavy quark symmetries and octet suppression to the
matrix element for $B\bar B$ mixing. Although the rates we compute are
too small to be observable at $B$-factories, they may be accessible to
experiments in hadronic colliders. We have demonstrated the method by
calculating the rate for $B^+\to D_{s,d}^{(*)+}e^+e^-$ to leading
order in the operator expansion and to leading-log order in
QCD. Systematic corrections to both of these expansions could and
should be computed. Irreducible errors are expected from the
application of the operator product expansion in the time-like regime,
which is analogous to the assumption of local quark-hadron duality in
calculations of heavy hadron lifetimes.

\bigskip

{\it Acknowledgments} This work is supported by the Department of
Energy under contract No.\ DOE-FG03-97ER40546.

\tighten

\end{document}